\documentclass[12pt]{article}

\usepackage{amsmath}
\usepackage{graphicx,psfrag,epsf,color}
\usepackage{enumerate}
\usepackage{natbib}
\usepackage{url} 

\usepackage{bm, bbm}
\usepackage{amssymb, amsfonts, amsthm,multirow,array,float,booktabs}

\newcommand{\blind}{0}

\begin{document}

\bibliographystyle{chicago}

\def\spacingset#1{\renewcommand{\baselinestretch}%
{#1}\small\normalsize} \spacingset{1}

\newcommand{\vect}[1]{\mbox{\boldmath $ #1$}}
\newcommand{\iid}{\stackrel{\mathrm{iid}}{\sim}}
\newcommand{\ind}{\stackrel{\mathrm{ind}}{\sim}}
\newcommand\Polya{P\'olya }

\newcommand{\K}{\mathcal{K}}


\if0\blind
{
  \title{\bf Mixture modeling on related samples by $\psi$-stick breaking and kernel perturbation}
  \author{Jacopo Soriano \thanks{Part of the research was completed while JS was a PhD student at Duke University.}\hspace{.2cm}\\
    Google Inc.\\\\
    Li Ma \thanks{LM's research is partly supported by NSF grant DMS-1612889 and a Google Faculty Research Award.}\hspace{.2cm}\\
    Duke University}
}
  \maketitle

\if1\blind
{
  \bigskip
  \bigskip
  \bigskip
  \begin{center}
    {\LARGE\bf Title}
\end{center}
  \medskip
} \fi
\vspace{-1.5em}

\bigskip
\begin{abstract}
There has been great interest recently in applying nonparametric kernel mixtures in a hierarchical manner to model multiple related data samples jointly. In such settings several data features are commonly present: (i) the related samples often share some, if not all, of the mixture components but with differing weights, (ii) only some, not all, of the mixture components vary across the samples, and (iii) often the shared mixture components across samples are not aligned perfectly in terms of their location and spread, but rather display small misalignments either due to systematic cross-sample difference or more often due to uncontrolled, extraneous causes. Properly incorporating these features in mixture modeling will enhance the efficiency of inference, whereas ignoring them not only reduces efficiency but can jeopardize the validity of the inference due to issues such as confounding. We introduce two techniques for incorporating these features in modeling related data samples using kernel mixtures. The first technique, called $\psi$-stick breaking, is a joint generative process for the mixing weights through the breaking of both a stick shared by all the samples for the components that do not vary in size across samples and an idiosyncratic stick for each sample for those components that do vary in size. The second technique is to imbue random perturbation into the kernels, thereby accounting for cross-sample misalignment. These techniques can be used either separately or together in both parametric and nonparametric kernel mixtures. We derive efficient Bayesian inference recipes based on MCMC sampling for models featuring these techniques, and illustrate their work through both simulated data and a real flow cytometry data set in prediction/estimation, cross-sample calibration, and testing multi-sample differences.
\end{abstract}

\noindent%
\vfill

\newpage
\spacingset{1.45} 
\section{Introduction}
\label{sec:intro}

Kernel mixtures are a powerful tool for modeling a variety of data sets, especially in the presence of a natural clustering structure \citep{escobar:1995,maceachern:1998}. A good portion of the rapidly expanding literature on Bayesian nonparametrics is aimed at building effective mixture models. A recent focus of the literature is on how to jointly model in a hierarchical manner data samples that are similar or otherwise related, the main objective being effective borrowing of strength across samples, thereby substantially enhancing inference on the underlying data generative mechanisms as well as prediction. This is particularly important for complex data sets, for which each individual sample may only contain very limited information regarding the underlying probability distribution. Among many notable efforts in this direction, \cite{lopes2003bayesian} proposed a hierarchical model for multiple finite mixtures. \cite{muller2004method} proposed a nonparametric extension of \cite{lopes2003bayesian}'s model by replacing finite mixtures with Dirichlet process (DP) mixtures. In a different vein, \cite{cron2013hierarchical} proposed to use the hierarchical DP, or HDP, \citep{teh2006hierarchical} as the mixing distribution to characterize variation across multiple mixture distributions. \cite{rodriguez2008nested} proposed the nested DP (NDP) mixture, which is an infinite mixture of DP mixtures that induces an additional level of clustering among multiple mixture distributions themselves (to be distinguished from the clustering within each mixture distribution).

While applicable to a variety of mixture modeling contexts, our work is motivated during our attempt to apply existing hierarchical mixture models to the analysis of data collected from flow cytometry experiments.  Flow cytometry is a laser-based technology that measures biomarkers on a large number of cells, so each cell is an observation from a distribution in $\mathbb{R}^p$, where $p$ is the number of biomarkers measured. The cell population typically comes from a blood sample in immunological studies, and it consists of cells of various subtypes---e.g., T cells, B cells, etc.---with each subtype forming a ``cluster'' in the sample space. Because each cell subtype has a specific function in the immune system, inference on the abundance of the various subtypes across blood samples of a patient under different stimulating conditions, for instance, is of interest. Mixture models are natural tools for characterizing such data as the data is indeed a mixture of various cell types \citep{chan2008statistical}, and because a typical flow cytometry study will involve multiple samples collected under different conditions, the need for jointly modeling to achieve effective borrowing of strength also naturally arises \citep{cron2013hierarchical}.

During the analysis of flow cytometry experiments using mixtures, we encountered a number of important challenges that we believe are present in numerous (if not most of) other applications involving mixture modeling of related samples (not only with location-scale kernels but beyond). Below we summarize the three main data features/challenges that motivate the current work:
\begin{itemize}
\item[I.] {\em  Samples often share clusters but with differing weights.} Related samples tend to share some (even most) of their clusters, and these common clusters vary across related samples in their weights. In flow cytometry, for instance, data samples often share a vast majority of the cell subtypes, and the most common type of variation across samples is the differences in the relative sizes of the subtypes. 
\item[II.] {\em Only some, not all, clusters vary.} Often, only a fraction, not all, of the clusters vary across samples. In flow cytometry, not all cell subtypes are affected by the experimental conditions of interest. Very often only one or two cell types are affected and thus vary across the samples while the rest do not.
\item[III.] {\em Misalignment across samples in shared clusters.} Even the same cluster shared among samples is often not perfectly aligned across samples, either due to actual systematic difference across the samples, or very often due to the presence of extraneous, uncontrolled additional sources of variation, i.e., some ``random'' effect. This is easily seen in mixtures of location-scale families, where the location and spread of some shared clusters differ to various extent across samples. Such misalignment is ubiquitous in flow cytometry data, with numerous potential causes. For example even tiny differences in the chemical concentrations applied in the experimental protocol across experiments can cause noticeable ``perturbations'' in the cell subtypes.
\end{itemize}
As far as we know, none of the existing hierarchical approaches satisfactorily address all of these issues in a single coherent framework. Table~\ref{tab:comparison_of_models} provides a summary of these data features and the extent to which some of the state-of-the-art methods (along with the method we propose herein) address each of them. 
\begin{table}[h]

\begin{center}
\resizebox{\textwidth}{!}{%
    \begin{tabular}{ c | c|c|c }
\hline\hline
      & Shared clusters & Only a subset  & Misalignment\\
 &  with varying weights & of clusters differ& in kernels\\
    \hline
\cite{lopes2003bayesian,muller2004method} &  Not allowed & Allowed  &Not allowed\\
\cite{teh2006hierarchical,cron2013hierarchical} & Allowed & Not allowed  &Not allowed\\
\cite{rodriguez2008nested} & Not allowed & Not allowed  &Not allowed\\
This work & Allowed & Allowed  & Allowed\\
  \hline\hline
    \end{tabular}}
\caption{Comparison of hierarchical mixture models in terms of how they cope with the three common data features/challenges in modeling multiple related data samples.}
\label{tab:comparison_of_models}
\end{center}
\end{table}

Specifically, the existing approaches exploit some aspects of these features but do not fully take them into account. By introducing a cluster-specific hierarchical relationship among the samples, \cite{lopes2003bayesian} and \cite{muller2004method} allow some clusters to be shared among the samples. However, their models require that the kernel parameters and the mixture weight for each cluster be either both shared across samples or both different, without the option to decouple these two different types of variations. In particular, no clusters are allowed to have only one type of variation---e.g., mixing weights---under these models. In the context of flow cytometry, for instance, this would mean that cell subtypes cannot change just in abundance across the samples but not in their location and spread, clearly an unrealistic assumption. On the other hand, by using the hierarchical DP \citep{teh2006hierarchical} as the mixing distribution, \cite{cron2013hierarchical} does allow variations to exist in weights alone, but enforces the constraint that all clusters must all vary across samples, excluding the common situation in applications such as flow cytometry that only some clusters (e.g., subtypes) vary while others remain unchanged across conditions. Finally, under the nested DP mixture \citep{rodriguez2008nested}, the clusters in each sample must either be completely identical as those in another sample if they fall into the same model level cluster or all be completely different, in both weights and kernel parameters, if they belong to different model level clusters.

New hierarchical modeling techniques are needed to address these limitations. To meet this need, we design two new modeling devices that can be embedded into a single hierarchical mixture modeling framework---the first for the mixing weights and the other for the kernel parameters. For the weights, we introduce a new stick breaking process that induces shared weights on some clusters (those that do not change in abundance) through breaking a ``shared'' stick across all samples while inducing different weights on the other clusters through breaking an ``idiosyncratic'' stick for each sample. This technique will allow us to address challenges~I and II. For the mixture kernels, we utilize a {\em hierarchical} kernel to induce local perturbations in the kernel parameters across samples, which mimics the effect on the kernels due to uncontrolled confounding. By decoupling the hierarchical relationship among the mixing weights from that among the kernel parameters, our approach offers the needed additional flexibility and thus achieves substantially higher efficiency in modeling related mixtures, as will be demonstrated through numerical examples. 

The rest of the paper is organized as follows. We start in Section~\ref{sec:background} with a brief review of the relevant background regarding nonparametric mixture modeling and stick breaking, and then in Section~\ref{sec:techniques} introduce the two  techniques in turn. In Section~\ref{sec:computation} we provide a recipe for posterior inference based on Markov chain Monte Carlo (MCMC) sampling. In Section \ref{sec:examples} we compare our method to current methods through simulation studies that cover prediction/estimation, cross-sample calibration, and testing multi-sample differences, and finally use it to analyze two flow cytometry data sets.

\section{Method}\label{sec:method}

\subsection{Background: Dirichlet process mixtures and stick breaking}
\label{sec:background}
While our techniques can be embedded into mixture models with various weight generating mechanisms and kernel families, we shall introduce and illustrate them in the context of DP mixtures of Gaussians, which is the most widely adopted nonparametric mixture model. 

Suppose $n$ observations $\vect{y}=(y_1,y_2,\ldots,y_n)$ are from a mixture model:
\begin{align*}
 y_i  \iid F, \quad i=1, \ldots, n, \quad \text{and} \quad f(\cdot)  = \sum_{k \in \mathcal{K}}  \pi_k \, g( \cdot| \lambda_k)
\end{align*}
where $f$ denotes the probability density function  of $F$, $g(\cdot | \lambda)$ is a kernel distribution parametrized by $\lambda$, $\pi_k$ the associated (mixture) weight, and $\mathcal{K}$ the countable (possibly infinite) index set of the mixture components (or clusters). Location-scale families are commonly adopted as the kernel distribution, in which case $\lambda_k$ specifies the location and spread of the $k$th cluster. By definition the weights satisfy $\pi_k \geq 0$ and $\sum_k \pi_k = 1$. An alternative and computationally attractive formulation utilizes a latent cluster membership label $Z_i\in\K$ for each observation, such that 
\[
 y_i\,|\,Z_i = k \sim g(\cdot|\lambda_k) \quad \text{and} \quad \Pr(Z_i = k) = \pi_{k} \quad \text{for $i=1,2,\ldots,N$ and $k\in\K$} .
\]

Bayesian inference under mixture models can proceed after specifying prior distributions on the weights and the
kernel parameters $\{(\pi_{k},\lambda_{k}):k\in\K\}$ \citep{marin:2005}.
A flexible and convenient choice on the prior for the mixing weights is a generative procedure called the stick breaking process (SBP) \citep{ayaram1994constructive,ishwaran2001gibbs}. The general scheme of SBP starts with the drawing of a sequence of independent random variables $v_1,v_2,\ldots$ supported on $(0,1)$. Then the weight for the $k$th cluster is given as 
\[ \pi_k = v_k\prod_{l=1}^{k-1} (1-v_l).\]
A popular two-parameter specification is the Poisson-Dirichlet process \citep{kingman1975random,pitman1997two},  corresponding to $v_i\sim {\rm Beta}(1-\gamma,\alpha+\gamma)$ for some parameters $\alpha$ and $\gamma$. In particular, when $\gamma=0$, this boils down to the weight generative mechanism from a Dirichlet process \citep{ferguson1973dp,ayaram1994constructive}, which we shall refer to as the SBP($\alpha$) process.

By adopting the SBP($\alpha$) prior on the weights, along with a prior $H$ on the kernel parameters, we obtain a Dirichlet process mixture (DPM) model:
\begin{align*}
\vect{\pi} = (\pi_k : k \in \mathcal{K}) & \sim \text{SBP}(\alpha) \quad \text{and} \quad \lambda_k \iid H, \quad k \in \K. 
\end{align*}
The most commonly adopted kernel distributions
are location-scale families such as the (multivariate) Gaussian family, i.e., $g(\cdot| \lambda_k) =
N(\cdot | \mu_k, \Sigma_k)$. In this case, $H$ is often chosen to be the corresponding conjugate prior such as  a normal-inverse-Wishart (NIW) prior on ($\mu_k,\Sigma_k$).

\subsection{Two techniques for hierarchically modeling related samples}
\label{sec:techniques}
Now assume $J$ samples of observations $\vect{y}_j = (y_{1,j}, \ldots,
y_{n_j,j})$ for $j=1, \ldots, J$ have been collected, and the observations in each sample are modeled by a mixture:
\begin{align*}
 y_{i,j} & \ind  F_j, \quad
i=1,
\ldots n_j\quad  \text{and} \quad j=1, \ldots, J \\
f_j(\cdot) & = \sum_{k \in \mathcal{K}} \pi_{j,k}\, g( \cdot| \lambda_{j,k}), \quad j = 1, \ldots, J,
\end{align*}
where $f_j$ is the probability density function of $F_j$, and $\lambda_{j,k}$ represent the kernel parameter for the $k$th cluster in the $j$th sample. To characterize potential relationship across the samples, let us assume that the $k$th component under each sample represent the same cluster (e.g., cell subtype). Note that this does not exclude the possibility of having novel clusters that appear in only one or some of the samples, in which case the weights $\pi_{j,k}=0$ if cluster $k$ is absent in the $j$th sample. Again we let $\K$ be the collection of all cluster indices over all the samples. Let $Z_{i,j}$ be a latent variable indicating that the data point $y_{i,j}$ belongs to the $k$th cluster with $k \in \mathcal{K}$. Then the model can be equivalently written as
\begin{align*}
[ y_{i,j} | Z_{i,j}  = k, \mu_{j,k}, \Sigma_{k} ] & \ind N(y_{i,j} |
\mu_{j,k}, \Sigma_{k}) \quad \text{and} \quad \Pr(Z_{i,j} = k) = \pi_{j,k} \text{ for $k\in\K$.}
\end{align*}

We next introduce techniques for prior choices on the weights and on the kernel parameters by extending the stick breaking prior and the kernel respectively, which will address the three data features and challenges described in the Introduction.

\paragraph{$\psi$-stick breaking for weights}

We consider a generative stick breaking procedure called ``$\psi$-stick breaking'' (for reasons to be explained below), which breaks $J$ sticks of unit length---one for each sample---in a dependent manner to generate the mixing weights $\{\pi_{j,k}:k=1,2,\ldots\}$ for $j=1,2,\ldots,J$. We start by observing that each cluster falls into one of two categories $\mathcal{K}_0$ and $\mathcal{K}_1$, that is $\K=\K_0\cup \K_1$ with $\K_0\cap\K_1 = \emptyset$: those in $\mathcal{K}_0$ have weights that do not vary across the $J$ samples (e.g., cell types whose abundance is constant across experimental conditions), i.e.,
 $ \pi_{j,k} =  \pi_{j', k}$  for $j,j'=1, \ldots, J$ for $k \in \mathcal{K}_0$, whereas those in $\mathcal{K}_1$ have varying weights across samples. 

\begin{figure}[t]
 \centering
 \includegraphics[width=0.75\textwidth]{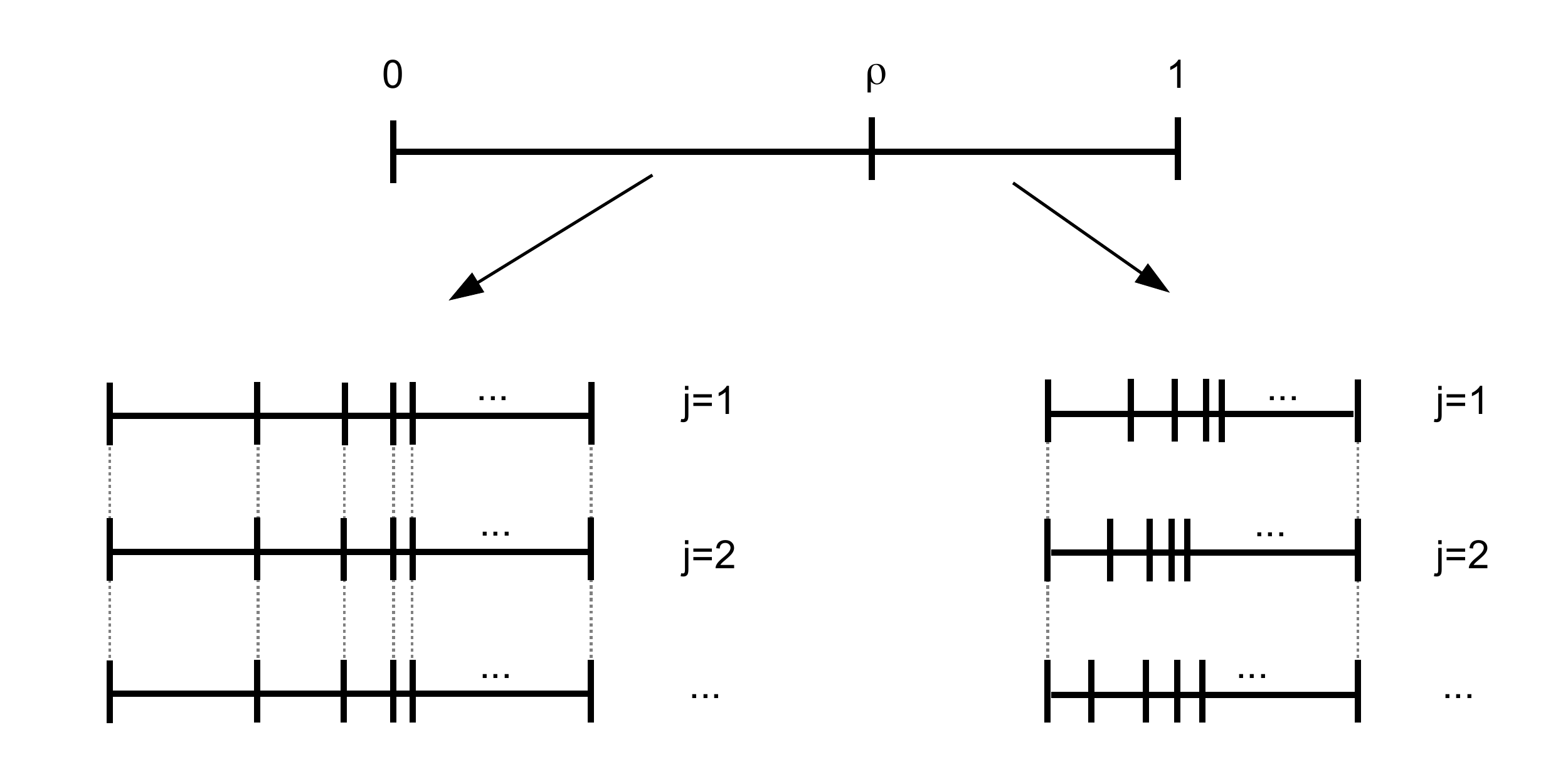}
 \caption{Illustration of the $\psi$-stick breaking procedure with the $s$-stick (left) and the $i$-sticks (right).}
 \label{fig:weights}
\end{figure}

The generative process proceeds in two steps and is illustrated in Figure~\ref{fig:weights}. In the first step, we break the $J$ sticks at exactly the same spot into two pieces of length $\rho$ and $1-\rho$ respectively, where $\rho\in (0,1)$ is drawn as a Beta random variable. 
Then in the second step, we use the $J$ pieces of length $\rho$ to generate the weights for the components in $\K_0$, and the $J$ pieces of length $1-\rho$ for the subtypes in $\K_1$.
Hence the parameter $\rho$ is interpreted as the overall proportion of the clusters with constant weights across samples. 

Specifically, one can imagine that we {\em tie} the $J$ sticks of length $\rho$ together and break them using a single SBP as if they were a single stick---always at the same locations. For this reason, we shall refer to the common stick formed by tying the $J$ sticks of length $\rho$ as the ``shared'' stick, or the $s$-stick. Let $\{w_{0,k}: k\in\K_0\}$ with $\sum_{k\in\K_0} w_{0,k}=1$ be the randomly generated {\em relative} sizes of the components in $\K_0$ in terms of the proportions of the $s$-stick. So the absolute size of each cluster that does not change across samples is given by $\pi_{j,k}=\rho w_{0,k}$ for all $j=1,2,\ldots,J$ and $k\in\K_0$.

On the other hand, we break the $J$ sticks of length $1-\rho$ {\em independently} using separate independent SBPs, each generating the weights for one of the $J$ samples, corresponding to the sizes of clusters that vary across samples. For this reason, we shall refer to the $J$ sticks of length $1-\rho$ as the ``idiosyncratic'' sticks, or the $i$-sticks. We let $\{w_{j,k}:k\in\K_1\}$ for $j=1,2,\ldots,J$ with $\sum_{k\in\K_1}w_{j,k}=1$ be the randomly generated lengths of the components as proportions of the corresponding $i$-stick. So for the $k$th cluster, its weight in the $j$th sample is given by $\pi_{j,k}=(1-\rho)w_{j,k}$. 

Using SBP$(\alpha)$ processes for breaking each of the $s$- and $i$-sticks, we arrive at a joint generative model for the weights in all of the $J$ samples, which we call ``shared/idiosyncratic" (si or $\psi$) stick breaking. Specifically, with a Beta prior on the length of the shared stick, we arrive at the following hierarchical model for weights 
\begin{align}\label{eq:infinite_prior_on_weights}
\pi_{j,k} & =
\left\{
\begin{array}{ll}
\rho w_{0,k} &  j=1, \ldots, J \text{ and } k \in \mathcal{K}_0 \\
(1 - \rho ) w_{j,k} & j=1, \ldots, J \text{ and } k \in \mathcal{K}_1
\end{array}
\right. \\
 \rho & \sim \text{Beta}(a_{\rho}, b_{\rho}) \nonumber \\
 (w_{0,k} :  k \in \mathcal{K}_0) & \sim \text{SBP}(\alpha) \nonumber \\
  (w_{j,k} : k \in \mathcal{K}_1) & \iid \text{SBP}(\alpha), 
\quad j=1, \ldots, J. \nonumber 
\end{align}
See Figure \ref{fig:weights} for a visualization of
the hierarchical prior on the mixture weights.

The hyperparameter $\alpha$  specifies the size of the clusters as well as the number of clusters (in $\K_0$ and $\K_1$  respectively), with a smaller $\alpha$ corresponding to a small number of large clusters and a larger $\alpha$ corresponding to a large number of small clusters. We infer on $\alpha$ in a hierarchical Bayesian paradigm by placing Gamma hyperprior on it:  
$ \alpha \sim \text{Gamma}(\tau_{\alpha,1},\tau_{\alpha,2} ) $.

\paragraph{Local kernel perturbation}

We utilize a hierarchical setup to incorporate local perturbation in the kernel parameters, thereby adjusting for the misalignment and allowing more effective borrowing of information across the samples on each cluster. Specifically, we model the kernel parameters $\{\lambda_{j,k}\}$ as follows
\begin{align*}
\lambda_{0,k} &\iid H_0(\cdot\,|\, \phi_0) \quad \text{for $k\in\K$}\\
\lambda_{j,k} &\iid H(\cdot\,|\,\lambda_{0,k},\epsilon) \quad \text{for $j=1,2,\ldots,J$}
\end{align*}
where $\lambda_{0,k}$ represent the cross-sample ``centroid'' kernel parameters for the $k$th cluster, with a hyperprior $H_0$ specified by hyperparameter $\phi_0$. Given $\lambda_{0,k}$, the sample-specific kernel parameters for the $k$th cluster $\lambda_{j,k}$ is drawn from $H$ with additional hyperparameter $\epsilon$, which specifies the dispersion of cluster $k$ among the samples around the ``centroid''. 

The above specification enforces that each cluster $k$ will have misalignment. More generally, in some problems misalignment may exist in only a subset of the clusters. To allow for such cases, again appeal to a ``spike-and-slab'' setup by introducing an additional Bernoulli latent indicator $S_{k}$ for each cluster, such that $S_k=1$ if there is misalignment in cluster $k$ whereas $S_k=0$ if otherwise. That is,
\[
\lambda_{j,k} \ind \begin{cases} \delta_{\lambda_0,k} & \text{if $S_{k}=0$}\\
H(\cdot|\lambda_{0,k},\epsilon) & \text{if $S_{k}=1$} 
\end{cases} \qquad \text{and} \qquad S_{k}\iid {\rm Bernoulli}(\varphi)
\]
where $\delta_{\cdot}$ represents a point mass.

\begin{figure}
 \centering
 \includegraphics[width=0.65\textwidth]{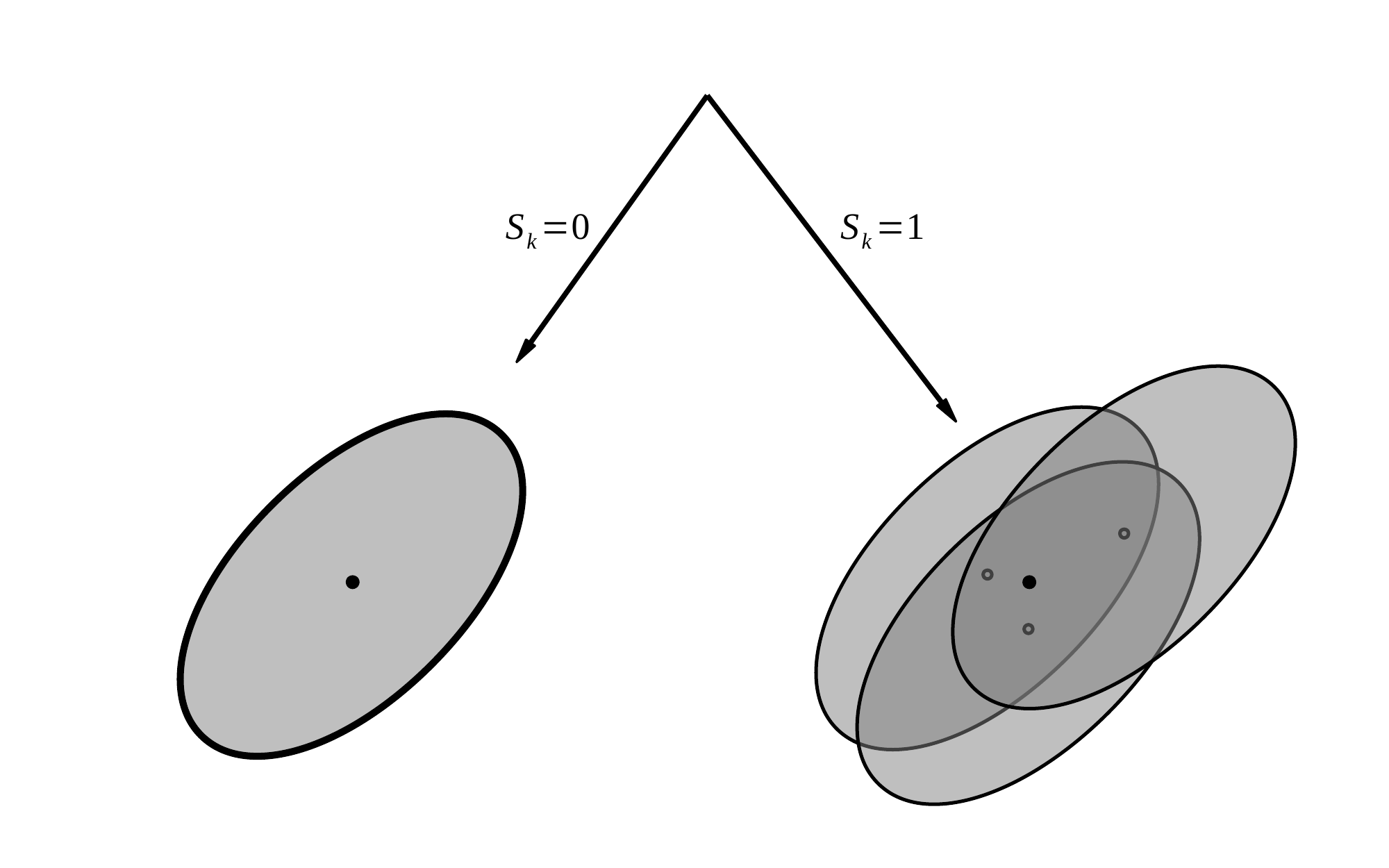}
 \vspace{-1em}
 \caption{A locally perturbed Gaussian kernel with a spike-and-slab setup. 
 When $S_k=0$, all kernels for the $k$th cluster are identical across samples. When $S_k=1$, the kernel is centered around a common mean but are not identical. }
 \label{fig:perturbation}
\end{figure}

Putting the pieces together in the context of Gaussian kernels, we arrive at the following spike-and-slab version of the locally perturbed kernel model:
\begin{align*}
\Sigma_k^{-1} & \iid \text{Wishart}(\Psi_1, \nu_1) \\
 [\mu_{j,k}| \mu_{0,k}, \Sigma_k, S_{k}] & \ind 
\delta_{\mu_{0,k}}  1_{\{ S_{k}=0\}}
+  \text{Normal}(\mu_{0,k},
\epsilon \Sigma_{k})  1_{\{ S_{k}=1\}} \\
[\mu_{0,k}|  \Sigma_{k}  ] & \ind \text{Normal}(m_1, \Sigma_{k}/ k_0) \\
S_{k} &\iid {\rm Bernoulli}(\varphi).
\end{align*}
This model is illustrated in Figure~\ref{fig:perturbation}. The hyperparameter $\epsilon$ specifies the
total amount of local variation between the means of each group $\mu_{j,k}$ and
the grand mean $\mu_{0,k}$, and $\varphi$ specifies the proportion of clusters that have misalignment.
The hyperparameters $m_1$, $\Psi_1$, $k_0$, $\epsilon$, and $\varphi$ are all characterizing ``global'' features of the data that pertain to all of the clusters and samples. We can reliably infer them by pooling information through hierarchical Bayes. In particular, in our numerical examples we adopt the following hyperpriors: 
$\epsilon \sim
\text{Uniform}(a_{\epsilon},
b_{\epsilon} )$, 
$ m_1 \sim \text{Normal}(  m_2 ,S_2)$, 
$  \Psi_1  \sim \text{Inverse-Wishart}(\Psi_2 ,\nu_2 )$, 
$ k_0  \sim \text{Gamma}( \tau_1/2,  \tau_2/2 )$, and $\varphi \sim \text{Beta}(a_{\varphi}, b_{\varphi})$.

\subsection{Posterior inference based on MCMC sampling}
\label{sec:computation}

Posterior inference can be carried out through Markov Chain Monte Carlo (MCMC).
One option is to use \cite{muller2004method}'s
standard \Polya urn scheme. A benefit of this sampling scheme is that all the random weights
are integrated out. However it can be computationally inefficient for large
datasets such as in flow cytometry experiments. 
Alternatively, one can approximate the nonparametric model with a finite model
and use a blocked Gibbs sampler \citep{ishwaran2001gibbs}, which is more
efficient in terms of mixing and computational speed, and hence is what we recommend.

To this end, two different finite approximation strategies are commonly adopted for DPMs and other stick breaking mixtures: (i) truncating the stick breaking at some maximum number of components and (ii) using finite-dimensional
symmetric Dirichlet distribution.  
These two approximations might look very different at first, but the
main difference between the two is in the induced stochastic ordering of the weights, which is
irrelevant in mixture models. 
In fact, as \cite{kurihara2007collapsed} points out, one can apply a size-biased permutation to the
order of the weights of a finite symmetric Dirichlet distribution and obtain a
distribution which is practically identical to the truncated SBP. However, the two strategies are not computationally equivalent for mixture models. The weights under the symmetric finite-Dirichlet approximation are
exchangeable, which
results in substantially improved mixing over truncating the SBP. Therefore we opt for the symmetric finite Dirichlet approximation in our implementation.
This approximation has been studied and used by many authors in a
variety of contexts. See \cite{neal2000markov}, 
\cite{green2001modelling} and \cite{ishwaran2002exact}, among others. 
Specifically, under this approximation, the infinite sequences of mixture weights in Eq.~\eqref{eq:infinite_prior_on_weights} are replaced by:
\begin{align*}
 (w_{0,k} :  k \in \mathcal{K}_0) & \sim \text{Dirichlet}(\alpha/K_0,\alpha/K_0,\ldots,\alpha/K_0) \\
  (w_{j,k} : k \in \mathcal{K}_1) & \iid\text{Dirichlet}(\alpha/K_1,\alpha/K_1,\ldots,\alpha/K_1),
\quad \text{for } j=1, \ldots, J,
\end{align*}
where $K_0$ and $K_1$ represent the numbers of mixture components that are
shared and differential across the groups, respectively.  
In the nonparametric case, both $\K_0$ and $\K_1$ are infinite, while in the finite
approximation we need to choose $K_0$ and $K_1$. 
A simple choice is to set $K_0 =  K_1 = K $ for some large $K$ which
represents an upperbound to the a priori expected number of mixture components.

With this specification, next we give the details on the MCMC sampler for the joint posterior in terms of the full conditionals:
\begin{enumerate}
 \item Latent assignments for  $ i = 1, \ldots, n_j $ and $ j=1, \ldots, J$:
$$
\Pr(Z_{i,j} = k |  \ldots ) \propto
\pi_{j,k} \text{Normal}(y_{i,j}|\mu_{j,k},
\Sigma_{k}), \quad  k \in
\mathcal{K}.
$$
\item Mixture weights:
\begin{align*}
[  w_{0,1},  \ldots, w_{0, K_0}  | \ldots ] & \sim
\text{Dirichlet}( n_{0,1} + \alpha/K_0, \ldots, n_{0,K_0 } +
\alpha/K_0) \\
[  w_{j,1},  \ldots, w_{j, K_1}  | \ldots ] & \ind
\text{Dirichlet}( n_{j,1} + \alpha/K_1, \ldots, n_{j,K_1 } +
\alpha/K_1),
\end{align*}
where $n_{0,k} = |Z_{i,j} = k\; : \; i=1, \ldots, n_j \; \text{and} \; j = 1,
\ldots, J |$ for $k \in \mathcal{K}_0$, and 
 $n_{j,k} = |Z_{i,j} = k \; : \; i=1, \ldots, n_j|$ for
$j=1, \ldots, J$ and $k \in \mathcal{K}_1$. 

\item Latent perturbation state variables for $k \in \mathcal{K}$:
$$
\Pr( S_k = 1 |  \ldots) = \bigg( 1 +
\dfrac{1-\varphi}{\varphi} \cdot
{\rm BF}_k \bigg)^{-1},
$$
where
\begin{align*}
 {\rm BF}_k & = \bigg( \dfrac{|\Psi_{1,k}^{(0)}| }{|\Psi_{1,k}^{(1)}| }
\bigg)^{(\nu_1+ \sum_j n_{j,k})/2} \prod_j  ( \epsilon n_{j,k} +
1 )^{p/2} \\
\Psi_{1,k}^{(1)} & = \bigg\{  \Psi_{1}^{-1} + \sum_j \bigg[ SS_{j,k}  +
\big(\epsilon + \dfrac{1}{n_{j,k}}\big)^{-1}
 (\bar{Y}_{j,k} - \mu_k) (\bar{Y}_{j,k} - \mu_k)' \bigg]   \bigg\}^{-1} \\
\Psi_{1,k}^{(0)} &= [ \Psi_1^{-1} +  SS_k  + \sum_j n_{j,k} (\bar{Y}_k - \mu_k)
(\bar{Y}_k -
\mu_k)'  ]^{-1},
\end{align*}
for $\bar{Y}_{j,k} = \sum_{i:Z_{i,j}=k} Y_{i,j} / n_{j,k}$,
$\bar{Y}_k = (\sum_{i,j:Z_{i,j}=k} Y_{i,j}) / (\sum_j n_{j,k})$, \\
 $SS_{j,k} = \sum_{
\{i : Z_{i,j}=k \}} (Y_{i,j} - \bar{Y}_{j,k})(Y_{i,j} -
\bar{Y}_{j,k})'$
and
 $SS_k =\sum_{ \{i,j : Z_{i,j}=k \}} (Y_{i,j} - \bar{Y}_k)(Y_{i,j} -
\bar{Y}_k)' $.
\item Precision matrices for $k \in \mathcal{K}$:
$$
[\Sigma_k^{-1} | \ldots   ]  \sim
\text{Wishart}\big(\Psi_{1,k}^{( S_k )},
\nu_1 + \sum_j{n_{j,k}} \big)
$$
\item Grand means for $k \in \mathcal{K}$:
$$
 [ \mu_k |  \ldots ] \sim
\text{Normal}\bigg(
m_{1,k}^{(S_k)}, \Sigma_k /(\sum_j (\epsilon S_k
+ 1/n_{j,k})^{-1}  + k_0 )  \bigg),
$$
\item Group means for $j=1, \ldots, J$ and $k \in \mathcal{K}$: 
\begin{align*}
 [ \mu_{j,k}| S_k = 0, \ldots ]  & \sim \delta_{\mu_k} \\  
 [ \mu_{j,k}| S_k = 1, \ldots   ]  & \sim
\text{Normal}\bigg( \dfrac{n_{j,k}
\bar{Y}_{j,k} +
 \mu_k/\epsilon}{n_{j,k} + 1/\epsilon} ,  \Sigma_k/(n_{j,k}+1/\epsilon) 
\bigg).
\end{align*}
\item A Metropolis step to explore different modes of the posterior distribution by swapping an index from $\mathcal{K}_0$ with an index from $\mathcal{K}_1$.
The proposal distribution is defined as follows. An initial index $k'$ is drawn
proportionally to $ \sqrt{ n_{j,k }}$ for $k \in \mathcal{K}$, where
$n_{j,k} = | (i,j) :  Z_{i,j} = k|$, and
a  second index $k''$ is drawn uniformly from $\mathcal{K}_0$ if $k' \in
\mathcal{K}_1$
and    
uniformly from $\mathcal{K}_1$ if $k' \in \mathcal{K}_0$. Since the proposal is
symmetric, the swap is accepted with probability:
$$
\min \bigg( \dfrac{{\rm E}_{ w,\rho}( \prod_{j,k} \pi_{j,k}^{n_{j,k}}|
\vect{Z}_{\text{new}} )}{{\rm E}_{ w,\rho}( \prod_{j,k} \pi_{j,k}^{n_{j,k}}|
\vect{Z})}, 1 \bigg),
$$
where $\vect{Z}$ and $\vect{Z}_{\text{new}}$ represent the vectors of the latent assignments
 before and after the swap. Since the mixture components are exchangeable within
$\mathcal{K}_0$ and $\mathcal{K}_1$, the acceptance probability 
depends only on the swapped indices. Similar strategies to improve the exploration of the
sample space have been proposed by  \cite{porteous2012gibbs} and
\cite{papaspiliopoulos2008retrospective}.

\item The Dirichlet pseudo-count parameter $\alpha$ is updated using a Metropolis-Hastings step with the following proposal:
$$
\alpha^* | \alpha \sim \text{Gamma}(\alpha ^ 2 \cdot a, \alpha \cdot a),
$$
where is $a$ is a tuning parameter calibrated in the burn-in. 
\item Mean shrinkage parameter
$$[k_0| \ldots] \sim \text{Gamma}( (\tau_1 + p \cdot K) / 2, (\tau_2 + \sum_k(
\mu_{0,k} - m_1)' \Sigma_k^{-1} (
\mu_{0,k} - m_1) ) / 2)$$

\item Variance parameter $[\Psi_1^{-1}| \ldots] \sim \text{Wishart}( (\Psi_2 + \sum_k \Sigma_k^{-1})^{-1}, K \cdot \nu_1 + \nu_2)$.
\item Centroid mean parameter $[m_1| \ldots] \sim {\rm Normal}(Vm, V)$,
where 
$$
m = S_2^{-1} m_2 + k_0 \sum_k  \Sigma_k^{-1} \mu_{0,k}
$$
and 
$$
V = ( S_2^{-1} + k_0\sum_k \Sigma_k^{-1} )^{-1}.
$$
\item The perturbation parameter $\epsilon$ is updated using a Metropolis step with the following proposal: 
$$
\text{Uniform}(a_{\epsilon},b_{\epsilon})
$$
\item The proportion of clusters with kernel misalignment $[\varphi | \ldots] \sim \text{Beta}(a_{\varphi} + s_0, b_{\varphi} + s_1)$,
where
$s_i = |S_k = i : k=1, \ldots, K|.$
\item The ``length'' of the shared stick $[\rho | \ldots] \sim \text{Beta}(a_{\rho} + n_0, b_{\rho} + \sum_j n_j)$,
where
$n_j = \sum_{k} n_{j,k}.$
\end{enumerate}

\section{Numerical examples}\label{sec:examples}

In this section we provide three numerical examples. In the first example
 data are simulated under different mixture distributions, and we  compare the goodness-of-fit of our method with respect to competing approaches. 
In the second example we illustrate through a simulated dataset how our model can be used to remove small distributional shifts across related mixture distributions. In the third example
  we  compare the performance
of our model to other competing methods in testing and identifying differences across distributions. 
In the fourth example we analyze two real flow cytometry datasets. In all of the examples, we shall refer to our Dirichlet process mixtures of Gaussians with $\psi$-stick breaking and kernel perturbation as CREMID, as it models Closely RElated MIxture Distributions.

\subsection{Example 1: Estimation and predictive performance}
In this first example, we investigate how CREMID helps achieve more effective borrowing of information across samples thereby enhancing predictive performance. To this end, we consider four simulation scenarios, representative of a vast variety of real applications. We use the sum of $L_1$ distances of the estimated univariate predictive densities from the true densities as measure of goodness of fit. (Note that we used this metric instead of the more natural log predictive score or the $L_1$ distance between the multivariate predictive density from the true density, because at the time of writing, the available software for the competitor HDPM provides the marginal predictive densities but not the other two metrics.)

We consider the following multi-sample scenarios in
$\mathbb{R}^4$. In each scenario, there are three data samples ($j=1,2,3$) and the sample size for each is 100.
Below we outline the four different scenarios. Some of the parameters are
omitted here, but provided in the Appendix.  
\begin{enumerate}
 \item Local shift:
 $$
  y_{i,j}| \vect{\mu}, \vect{\Sigma}, \vect{\pi}  \sim 
   \pi_1
N(y_{i,j}| \mu_{1} + \delta_j, \Sigma_{1}) +  \sum_{k=2}^4 \pi_k
N(y_{i,j}| \mu_{k}, \Sigma_{k}), 
 $$
where $\delta_j =  ( j/2, 0, 0, 0)$ and $\mu_k \sim U(0,
10)$ for
$k=1, \ldots, 4$.

 \item Global shifts:
 $$
  y_{i,j}| \vect{\mu}, \vect{\Sigma}, \vect{\pi}  \sim 
     \sum_{k=1}^4 \pi_k
N(y_{i,j}| \mu_{k} +  \dfrac{j}{10} \mathbbm{1}_4, \Sigma_{k}), 
 $$
where  $\mu_k \sim U(0, 10)$ for
$k=1, \ldots, 4$.

 \item Local weight difference:
\begin{multline}
  y_{i,j}| \vect{\mu}, \vect{\Sigma}, \vect{\pi}  \sim 
   (\pi_{1} - 0.04(j-1)) N(y_{i,j}| \mu_{1}, \Sigma_{1}) \\
   + (\pi_{2} + 0.04(j-1)) N(y_{i,j}| \mu_{2}, \Sigma_{2}) 
   +  \sum_{k=3}^4 \pi_k
N(y_{i,j}| \mu_{k}, \Sigma_{k}), 
\end{multline}
where $\vect{\pi} = (0.09, 0.01, 0.8, 0.1)$ and $\mu_k \sim U(0, 10)$ for
$k=1, \ldots, 4$.

 \item Global weight differences:
\begin{align*}
  y_{i,j}| \vect{\mu}, \vect{\Sigma}, \vect{\pi}  & \sim 
        \sum_{k=1}^8 \pi_{j,k}
N(y_{i,j}| \mu_{k}, \Sigma_{k}) \\
\pi_{j}  & \propto \exp(m_j) \\
m_{j} & \sim N(0, S), 
 \end{align*}
where $\mu_k \sim U(0, 10)$ for
$k=1, \ldots, 8$.
\end{enumerate}

We compare our method to 
 \cite{muller2004method}'s hierarchical Dirichlet process mixture (HDPM) method. We use the  R package \texttt{DPpackage} \cite[]{jara2011dppackage} for fitting 
HDPM. In addition, we also compare these to methods to independent finite mixture of Gaussians for each of the three samples, using Mclust \citep{fraley:2002}, available in the R package \texttt{mclust}.
 
 \begin{figure}[h]
 \centering
 \includegraphics[width=0.8\textwidth]{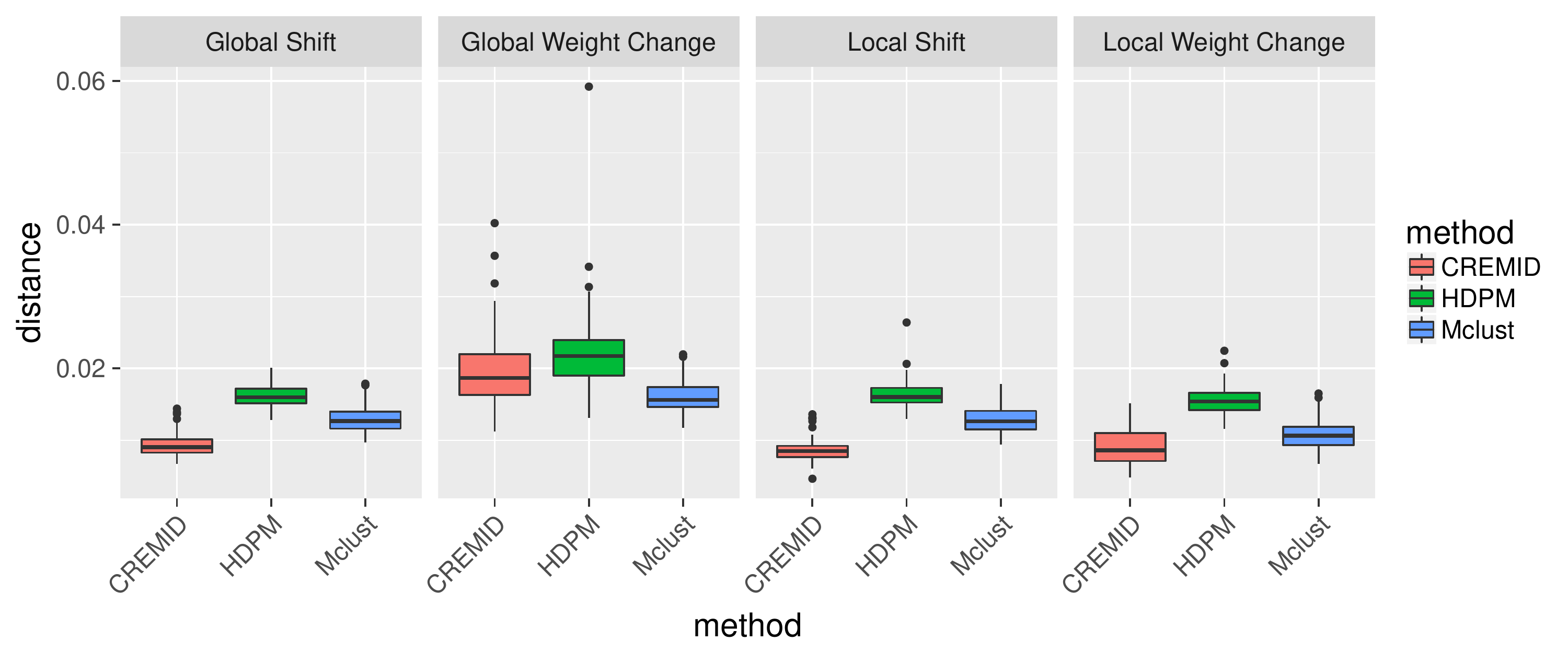}
 \caption{Box-plots of the sum of $L_1$ distances of the estimated univariate predictive densities from the true densities for three methods.}
 \label{fig:distance}
\end{figure}
 
 In Figure \ref{fig:distance} we show the sum of $L_1$ distances of the estimated univariate predictive densities from the true densities for the three methods. Our approaches outperform HDPM and mclust in the two shift scenarios. CREMID is the most accurate method in the two location shift scenarios as well as in the local weight change scenario. In the global weight change scenario, both our method and HDPM underperforms Mclust. Because the samples are different in all cluster weights, we pay a price for assuming that some cluster weights are shared.

\subsection{Example 2: Correcting for cross-sample misalignment}
 A common problem in studies involving data collected from multiple labs or centers is the misalignment of the same clusters across samples due to external confounders, which is what motivated our hierarchical locally perturbed kernel construction. In flow cytometry, for example, misalignment across cell subpopulations can be substantial. An important preprocessing step is cross-sample calibration---that is, to estimate and correct for the misalignment across samples and thereby produce ``standardized'' data sets for follow up studies. (This shares the registration problem in functional data analysis.) To this end, we note that for each observation $y_{i,j}$, if $Z_{i,j}=k$, that is, the observation belongs to cluster $k$, then we can compute a corrected value by adjusting for the shift in the cluster center across the samples:
\[ \tilde{y}_{i,j} = \mu_{0,k} +(y_{i,j} - \mu_{j,k}) = y_{i,j} - \Delta_{j,k}\]
where $\Delta_{j,k}=\mu_{j,k}-\mu_{0,k}$ is the displacement of cluster $k$ in sample $j$ relative to the centroid. Because $Z_{i,j}$ is unobserved, we can appeal to Bayesian model averaging (BMA) by computing the posterior mean of $\tilde{y}_{i,j}$
\[
{\rm E}(\tilde{y}_{i,j}\,|\,\vect{y}) 
= y_{i,j} - {\rm E}(\Delta_{j,Z_{i,j}}\,|\,\vect{y})\approx y_{i,j} - \dfrac{1}{B} \sum_{b=1}^B \Delta_{j,Z_{i,j}^{(b)}}^{(b)},
\]
where $\Delta_{j,Z_{i,j}^{(b)}}^{(b)}$ is the $b$th posterior draw on the displacement $\Delta_{j,Z_{i,j}^{(b)}}^{(b)}=\mu_{j,Z_{i,j}^{(b)}}^{(b)}-\mu_{0,Z_{i,j}^{(b)}}^{(b)}$. 

\begin{figure}[h]
 \centering
 \includegraphics[width=0.9\textwidth]{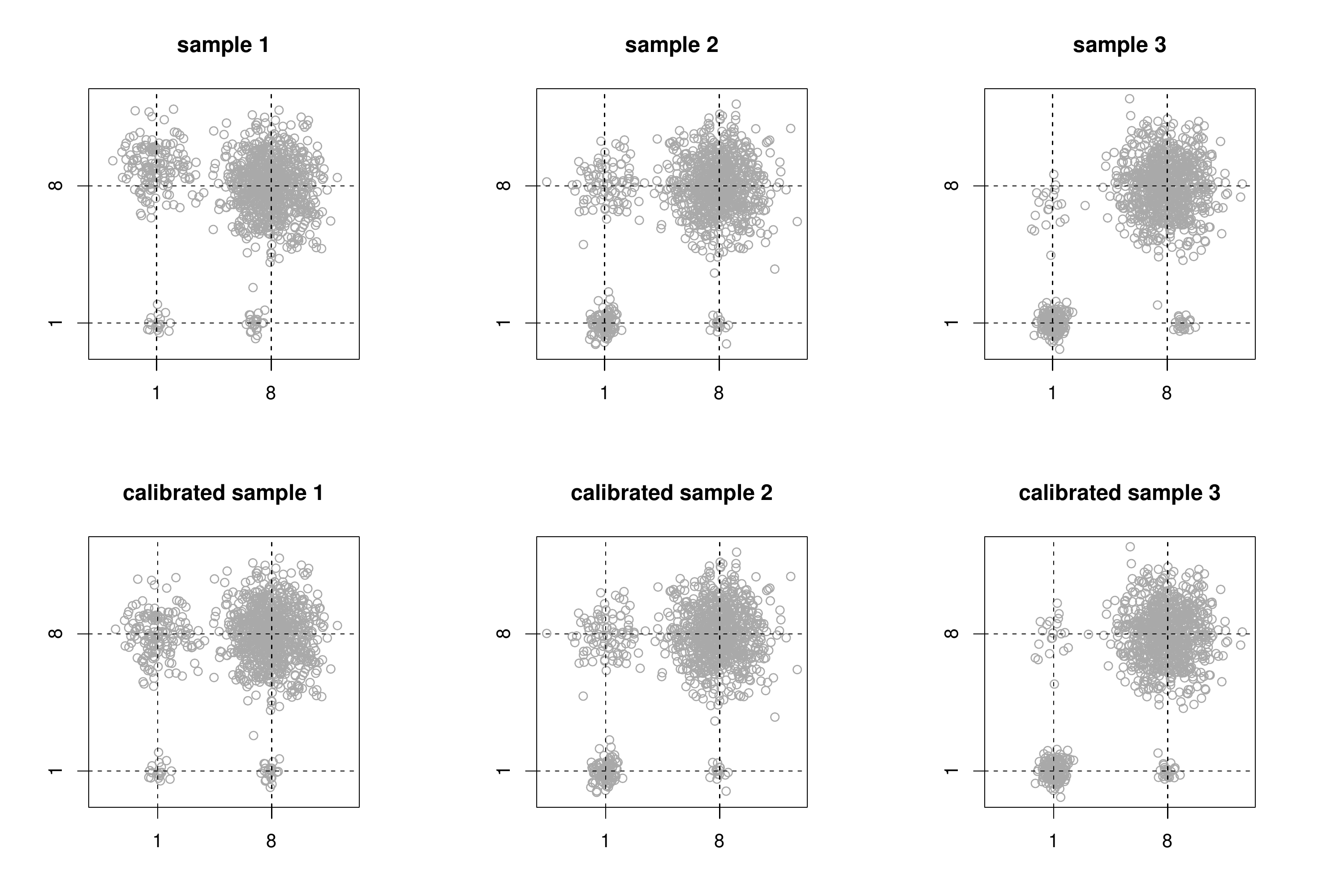}
 \caption{ The three plots in the first row show the data from Example 2
projected along the first two dimensions for each of the three samples. 
In the second row the three plots show the calibrated data, i.e., after removing the estimated kernel perturbations.}
 \label{fig:2}
\end{figure}

Let us consider a numerical example  based on 
mixture of normals in $\mathbb{R}^4$ to illustrate how one can remove cross-sample misalignment.
The data are generated as follows:
\begin{align*}
 y_{i,1} & \sim  0.16 N(\mu_{1,1}, I) + 0.80 N(\mu_{2}, 2I) +
 0.02 
N(\mu_{3},0.2 I) + 0.02 N(\mu_{1,4}, 0.1 I) \\
 y_{i,2} & \sim  0.09 N(\mu_{2,1}, I) + 0.80 N(\mu_{2}, 2I) + 
 0.09
N(\mu_{3},0.2 I) + 0.02 N(\mu_{2,4}, 0.1 I) \\
 y_{i,3} & \sim  0.02  N(\mu_{3,1}, I) + 0.80 N(\mu_{2}, 2I) + 
 0.16 
N(\mu_{3},0.2 I) + 0.02 N(\mu_{3,4}, 0.1 I),
\end{align*}
where $i=1, \ldots, 1000$, $\mu_{j,1}  =  (1,10-j,1,9)$, $\mu_2 = (8,8,8,8)$,
$\mu_3=(1,1,1,1)$ and
$\mu_{j,4} = (6+j,j,7,1)$.
 The three plots in the first row of Figure \ref{fig:2} show the data projected
along the first two dimensions for each of the three distributions. Most of the data ($80\%$) belong to a mixture component which is identical across the three distributions. The remaining $20\%$ of the data belong to three mixture components which are different across the three distributions. The means of two mixture components are shifted across the three distributions, while two mixture components have different abundance across the three distributions. 
The dashed lines in the plots help the reader identifying the across-sample shift in the means.

In the second row of Figure \ref{fig:2} the three plots show the calibrated data, i.e., after removing the estimated kernel perturbations. The model is able to correctly remove the local distributional shifts across the samples. 

\subsection{Example 3: Testing cross-sample differences in cluster weights}
\label{ex:cross_sample}
\begin{figure}[ht]
 \centering
 \includegraphics[width=0.6\textwidth]{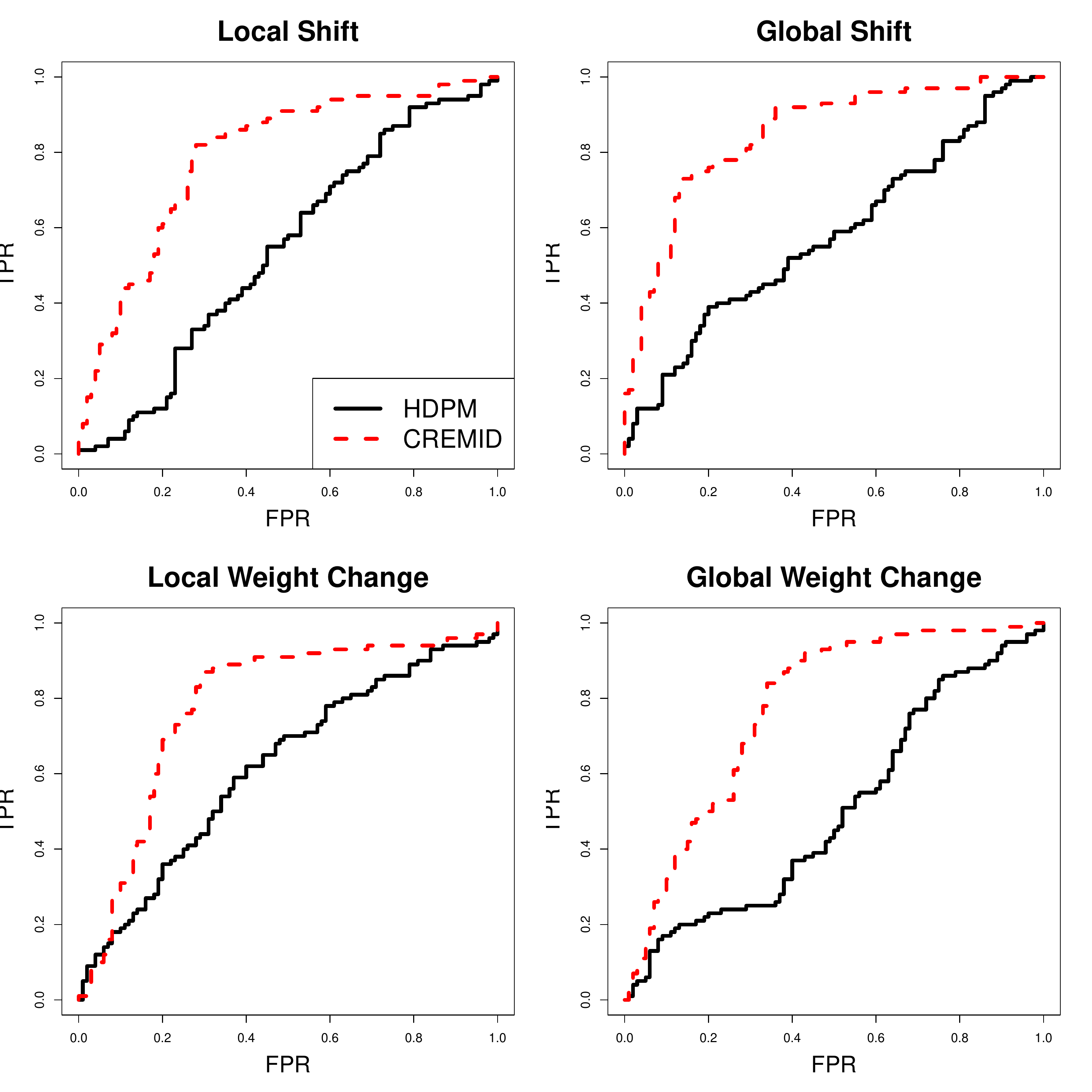}
 \caption{
ROC curves for two methods in Example~\ref{ex:cross_sample}: HDPM \citep{muller2004method} in black solid, our method in red dashed.}
 \label{fig:1}
\end{figure}

We consider the same multi-sample scenarios in
$\mathbb{R}^4$ used in Example~1. For each dataset we define a corresponding \emph{null} data set by permuting the labels of the three samples. In Figure \ref{fig:1} we compare the ROC curves of our method and HDPM
for testing the hypothesis that the three distributions are identical. Our method is substantially more powerful than HDPM in all four scenarios. 

In these simulations, for our method we use ${\rm E}( \rho \varphi  |\vect{y})$ as the test statistic.
This quantity goes to zero when there are differences in the mixture weights or in the mixture kernels across samples, and it goes to one when the distributions are identical across samples. One can adopt different test statistics under our method depending on the inference objective. For instance, if one is interested in testing just the presence of differences in weights then a suitable test statistic is ${\rm E}(\rho |\vect{y})$.

We compare our method only to HDPM since Mclust does not provide a way to test for differences across samples.
In HDPM  each $F_j$ is defined as a mixture of two
components: $F_j = \epsilon H_0 + (1-\epsilon) H_j$ for
$j=1, \ldots, J$. The distribution $H_0$
represents the common part, and $H_j$  represents the idiosyncratic part.
The hyperparameter $\epsilon$ controlling the
``degree of similarity'' across the $F_j$'s has a beta hyperprior. We use
${\rm E}(\epsilon|\vect{y})$ as the test statistic. 

\subsection{Application: flow cytometry}
In flow cytometry experiments, biomarkers are measured on a large
number of blood cells.  Different cell subtypes, i.e., groups of cells sharing
similar biomarker's levels, have distinct functions in human immune system.
Identifying variations in the abundance of subtypes across multiple samples
is an important immunological question. 
Additionally, the location of
a given subtype across samples can slightly change due to both experimental
variability and other uncontrolled ``random effects''.

We analyze two datasets where each one contains three samples of 5,000 blood
cells, and for each cell six biomarkers have been measured.  

\subsubsection{A control study}
 The blood from a given patient was split in three samples, and each
sample went through a separate experimental procedure to generate the data. Since the three samples are essentially biologically identical, one expects
no variations in the abundance of the different subtypes or large location shifts of the cell types. Small perturbations of the cell types are likely due to additional variations in the experimental procedures.

\begin{figure}
 \centering
 \includegraphics[width=\textwidth]{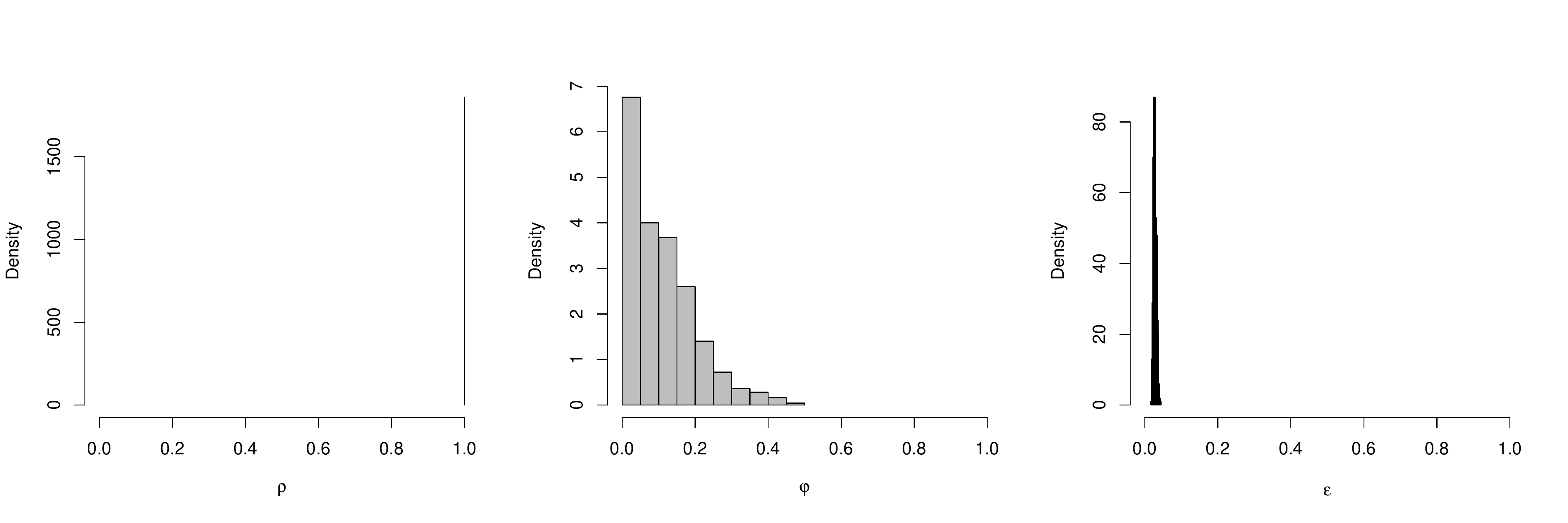}
 \caption{Histograms of the posterior of $\rho$ and $\epsilon$ for the flow cytometry control study. }
 \label{fig:2a_1}
\end{figure}

In Figure \ref{fig:2a_1} we plot the posterior distributions of $\rho$ and $\epsilon$ for this data set under our proposed model. 
The parameter $\rho$ reflects the total mass
assigned to mixture components where the mixture weights are identical across 
groups. 
In this dataset {\em a posteriori} this parameter concentrates around one, indicating that there is no
evidence of a difference in the mixture
weights across the three replicates.
The parameter $\epsilon$ controls the expected amount of shift in the location of each kernel across samples.
Its posterior does not
 concentrate around zero, indicating the presence of small misalignment among the replicate samples due to uncontrolled sources of variation. 
It is the decoupling of these two sources
of variations that allows us to correctly infer the absence of variations in the
mixture weights across the distributions of the three samples.

\subsubsection{Samples under different stimulation conditions}

\begin{figure}
 \centering
 \includegraphics[width=\textwidth]{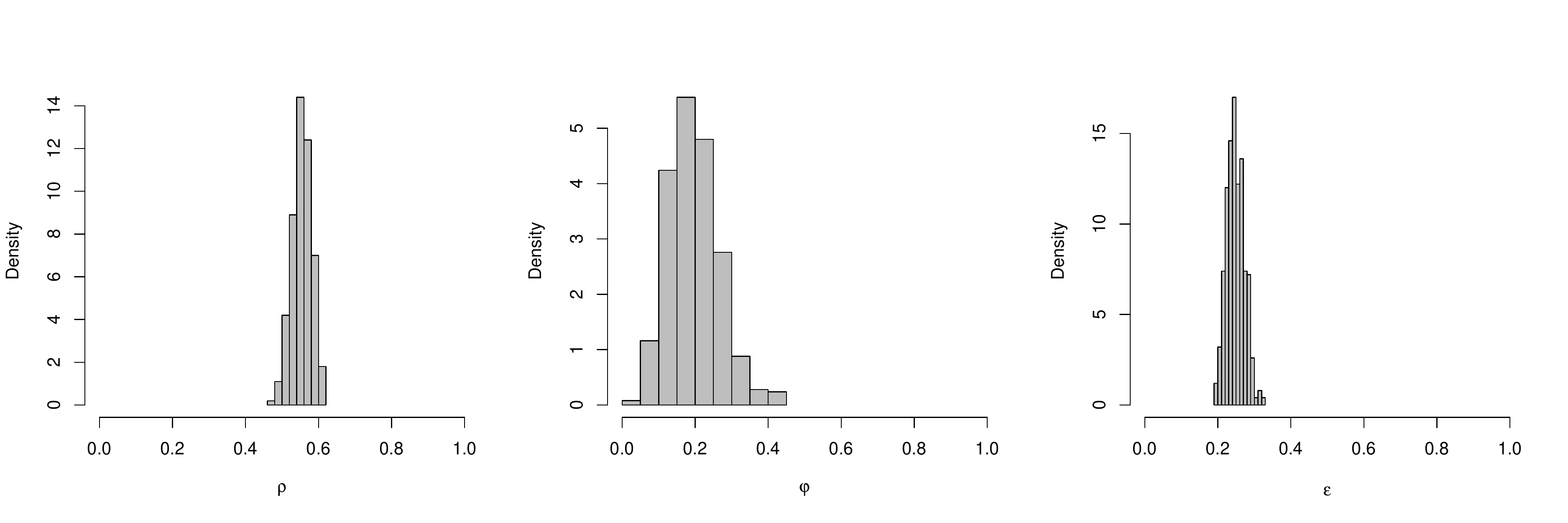}
 \caption{Histograms of the posterior distributions of $\rho$ and $\epsilon$. }
 \label{fig:3a_1}
\end{figure}

In another data set, three blood samples from an individual underwent different stimulation treatments. One sample was left unstimulated, while the two remaining samples were stimulated with CEF and CMV pp65, respectively. The samples underwent separate experimental procedures in data generation. In Figure \ref{fig:3a_1} we plot the posterior distributions of $\rho$
and $\epsilon$. The parameter $\rho$ concentrates
around 0.6, indicating that there are differences in some of the mixture
weights across the three samples.
The parameter $\epsilon$ concentrates around $0.2$, either due to effects of the experiment conditions on the locations of the kernels, which is also a systematic cross-sample difference, or substantial additional variations in the experimental procedures in comparison to the control study.  

To judge the goodness-of-fit, we also compare the predictive performance of our model with Mclust, evaluated by the log predictive likelihood of the a ``test'' sample. We randomly select 1,000 data points from the whole data set as a ``test'' sample, while using 5,000 observations as the ``training sample''. We had hoped to compare our method to other methods such as \cite{muller2004method} but at the time of writing, the existing software in {\tt R} (the {\tt HDPMdensity} function in {\tt DPpackage}) crashes for the data sets, most probably due to the large sample sizes, and it does not output predictive scores. 

\begin{table}[h]
\begin{center}
\begin{tabular}{lcc}
\hline\hline
&\multicolumn{2}{c}{Method} \\
\cmidrule{2-3}
Data set & CREMID & MClust \\
\hline
Control study&-15456.34&-16310.93\\
Different stimulation conditions &-14649.47&-15408.23\\
\hline

\hline
\end{tabular}
\end{center}
\caption{Log-$p$ predictive score comparison for CREMID versus MClust. Larger values (or smaller absolute values for negative scores) indicate better fit to the data.}  
\label{tab:log_predictive_flow}
\end{table}

\section{Conclusion}
\label{sec:conc}

In this work we have introduced two useful techniques in modeling related data sets using mixture models---the shared-idiosyncratic stick breaking and the locally perturbed kernel. When used together, they incorporate three common data features observed in real applications---(i) samples often share the same clusters with different weights; (ii) only some clusters vary across samples; (iii) misalignment in the clusters due to extraneous causes. We have derived Bayesian inference recipe through MCMC sampling and carried out an extensive numerical studies to illustrate the gain in inferential efficiency in both estimation, prediction, and hypothesis testing.

Finally, we note that while the two techniques are introduced and demonstrated in the context of mixtures of location-scale families, they are generally applicable to modeling related mixtures of other forms of kernels as well, such as mixtures of generalized linear models and mixtures of factor models. The computational details will vary but the general ideas remain the same.

\section*{Software}
R code for the proposed MCMC sampler and code for the numerical examples are available at \url{https://github.com/jacsor/cremid/} and 
\url{https://github.com/jacsor/MPG-examples/}, respectively.

\section*{Acknowledgment}
The authors are very grateful to Cliburn Chan for helpful discussions. The flow cytometry data set was provided by EQAPOL (HHSN272201000045C), an
NIH/NIAID/DAIDS-sponsored, international resource that supports the development, implementation, and oversight of quality assurance programs (Sanchez PMC4138253).

\section*{Appendix}

\subsection*{Numerical Examples}

\begin{enumerate}
 \item Local and global shift scenarios:
 \begin{align*}
  \Sigma_1(i,i) & = 1.1 \quad \text{for } i=1, \ldots, 4, \quad   \Sigma_1(i,j)
 = 0.9 \quad \text{for } i\neq j \text{ and } i,j=1, \ldots, 4; \\
  \Sigma_2(i,i) & = 2.0 \quad \text{for } i=1, \ldots, 4, \quad   \Sigma_2(i,j)
 = 1.0 \quad \text{for } i\neq j \text{ and } i,j=1, \ldots, 4; \\
  \Sigma_3(i,i) & = 0.4 \quad \text{for } i=1, \ldots, 4, \quad   \Sigma_3(i,j)
 = -0.1 \quad \text{for } i\neq j \text{ and } i,j=1, \ldots, 4; \\
  \Sigma_4(i,i) & = 0.1 \quad \text{for } i=1, \ldots, 4, \quad   \Sigma_4(i,j)
 = 0.0 \quad \text{for } i\neq j \text{ and } i,j=1, \ldots, 4; \\
 \vect{\pi} & = (0.3, 0.3, 0.2, 0.2).
 \end{align*}
 
  \item Local weight difference:
  $\Sigma_k$ for $k=1, \ldots, 4$ are identical to the local shift scenario
  and the global shift scenario.
  
  \item Global weight differences:
  \begin{align*}
   \Sigma_1 & = \text{diag}(1,1,1,1); \\
   \Sigma_2 & = \text{diag}(2,2,2,2); \\
   \Sigma_3 & = \text{diag}(0.2,0.2,0.2,0.2); \\
    \Sigma_k & = \text{diag}(0.1,0.1,0.1,0.1) \quad \text{for } k = 4, \ldots,
8.   
  \end{align*}
\end{enumerate}

\bibliographystyle{Chicago}

\bibliography{Bibliography}
\end{document}